\documentclass[12pt]{article}

\makeatletter

\@addtoreset{equation}{section}
\makeatother

\usepackage{amssymb}
\usepackage{amscd}
\usepackage{bm}
\usepackage{wick}
\usepackage{cite}

\bmdefine{\bx}{x}
\bmdefine{\by}{y}
\bmdefine{\bk}{k}
\bmdefine{\bp}{p}
\bmdefine{\bq}{q}

\begin{document}

\begin{titlepage}
\setcounter{page}{0}
\begin{flushright}
\begin{tabular}{l}
	hep-th/0606183\\
	KEK-TH-1092\\
\end{tabular}
\end{flushright}
\vskip 2.0cm

\begin{center}

	{\LARGE\bf Noncommutative Quantization \\[12pt]
	for Noncommutative Field Theory}

\vspace{15mm}

\renewcommand{\thefootnote}{\fnsymbol{footnote}}
	{\large Yasumi Abe\footnote[1]{email:\ yasumi@post.kek.jp}}\\[10mm]
	\noindent {\em Institute of Particle and Nuclear Studies\\
	High Energy Accelerator Research Organization (KEK)\\
	Tsukuba 305-0801, Japan }\\
\vspace{20mm}

\begin{abstract}
	We present a new procedure for quantizing field theory models on a noncommutative spacetime.
	The new quantization depends on the noncommutative parameter explicitly and reduces to the canonical quantization in the commutative limit.
	It is shown that a quantum field theory constructed by the new quantization yeilds exactly the same correlation functions as those of the commutative field theory, that is, the noncommutative effects disappear completely after quantization.
	This implies, for instance, that by using the new quantization, the noncommutativity can be incorporated in the process of quantization, rahter than in the action as  conventionally done.
\end{abstract}

\end{center}
\end{titlepage}

\section{Introduction}\mbox{}

	Noncommutative field theory has recently been attracting a lot of interests since the seminal work of Seiberg and Witten \cite{Seiberg:1999vs} which showed that a particular low energy limit of string theory is described by a gauge theory on a noncommutative space defined by the commutation relation for coordinates:
\begin{equation}
	[x^\mu,x^\nu]=i\theta^{\mu\nu},\label{NC space}
\end{equation}
	where $\theta^{\mu\nu}$ is an antisymmetric, constant matrix.
	The noncommutative relation (\ref{NC space}) suggests that we should treat coordinates, or fields depending on them, as operators rather than c-numbar quantities.
	However, if we introduce the Moyal star product
\begin{equation}
	f(x)\ast g(x)=\exp\Big(\frac{i}{2}\theta^{\mu\nu}\partial_\mu\partial'_\nu\Big)f(x)g(x')\Big\vert_{x'\rightarrow x}, \label{ast product}
\end{equation}
	and replace all the products between fields by this star product, we can still treat fields on a noncommutative space as c-number functions.
	Indeed, the Moyal star product is a noncommutative product and reproduces the commutation relation (\ref{NC space}):
\begin{equation}
	[x^\mu,x^\nu]_\ast=x^\mu\ast x^\nu-x^\nu\ast x^\mu=i\theta^{\mu\nu}.
\end{equation}
	The conventional prescription to define the action for a field theory on the noncommutative space is to replace all the products in the action for an ordinary commutative field theory by the Moyal star product (\ref{ast product}).
	Quantized versions of noncommutative fied theories are also considered, and there have been a large number of papers investigating their properties such as UV/IR mixing, nonrenormalizability, noncommutative standard model, violation of Lorentz invariance and the change of the causal structure (for reviews, see, for example, \cite{Douglas:2001ba,Szabo:2001kg,Konechny:2000dp} and, for phenomenological aspects, see also \cite{Hinchliffe:2002km, Hewett:2000zp}).

	In all such studies on noncommutative quantum field theory (QFT), the field is quantized based on the canonical commutation relation.
	The basic reason for this seems that the canonical quantization gives an accurate description of nature in low energy scales where experimental data are available.
	However, there is no evidence to believe that the canonical quantization is the correct prescription up to high energy scales such where the noncommutative field theory would be realized.
	In principle, in such high energy scales, there could be an exotic quantization, other than the canonical quantization, which has the same low energy limit as that of the canonical quantization.

	In this paper, we show that such a quantization does exist by the example of noncommutative scalar field theories.
	This new quantization, which we call {\it noncommutative quantization}, depends on the noncommutative parameter $\theta^{\mu\nu}$ such that it reduces to the standard canonical quantization when $\theta^{\mu\nu}\rightarrow0$.
	We shall see that Green's functions and the S-matrix elements appropriate for the new quantization can be defined and calculated by means of perturbative expansions in the same way as an ordinary QFT.
	 To our surprise, they turn out to be equivalent to those of the corresponding commutative QFT in all orders of perturbation.
	In fact, this equivalence holds even nonperturbatively, and we can find an exact map from a commutative QFT to our new QFT.
	Interestingly, the new QFT yeids exactly the same dynamics as those of the commutative field theory, even though the action includes noncommutative interaction terms explicitly.
	This is because the noncommutativity in interaction terms and the noncommutativity introduced in the quantization procedure cancel out each other.
	This peculiar structure of new quantization implies, for instance, that if we start from an action of a commutative field theory and quantize it by the noncommutative quantization, then we get the dynamics equivalent to a noncommutative field theory quantized by the canonical quantization.
	We will also investigate the structure of this cancellation of the noncommutativity in detail.

	This paper is organized as follows.
	In section 2, we present the noncommutative quantization for noncommutative scalar field theories with polynomial interaction terms.
	In section 3, this new quantization is applied to a free field.
	It is shown that operators similar to the laddar operators can be introduced, with which a ``Fock space" representation is constructed.
	Section 4 is devoted to the investigation of the noncommutative quantization for an interacting field theory.
	Green's functions and the S-matrix are defined and their equivalence to those of the commutative QFT is discussed.
	The structure of this equivalence is argued in some detail in section 5.
	Our conclusion and discussion are given in section 6.

\vspace{9mm}

\section{Noncommutative Quantization}\mbox{}

	To present our new quantizaiton for field theory models on a noncommutative space with Lorentzian metric $(+,-,-,\cdots,-)$, we consider, for definitness, we will consider only $d+1$ dimensional real scalar field theories whose interaction terms are given by the (noncommutative) polynomial form:
\begin{equation}
 \begin{array}{rcl}
	\mathcal{L}&=&{\displaystyle \frac{1}{2}\Big[(\partial_\mu\phi)^2-m^2\phi^2\Big]+\sum_{n=3}^\infty\frac{\lambda_n}{n!}\overbrace{\phi\ast\phi\ast\cdots\ast\phi}^n}\\
	&=&\mathcal{L}_0+\mathcal{L}_\textrm{\scriptsize{int}}.\rule{0mm}{7mm}
 \end{array}
	\label{lagrangian}
\end{equation}
	For example, the ordinary noncommutative $\phi^4$ theory is given by the case where only $\lambda_4$ is nonzero.

	To proceed further, we impose time-space commutativity condition, $\theta^{0i}=0$, so that we are allowed to define the canonical momentum and the Hamiltonian by the conventional formulae\footnote{
	This is because, as we shall see bellow, we deform the canoncal quantization to define a new quantization procedure.
	If $\theta^{0i}\ne0$, canonical formalism would become somewhat complex since a interaction term of an action contains any orders of time derivatives\cite{Gomis:2000gy}, and then our new quantization would also become complex.
	For easiness, we will not consider such cases in this paper.
}:
\begin{equation}
	\pi(x)=\frac{\delta S}{\delta \dot{\phi}(x)}=\dot{\phi}(x),\label{canonical momentum}
\end{equation}
\begin{equation}
	H=\int d^dx\;\Big[\dot{\phi}(x)\pi(x)-\mathcal{L}\Big],\label{hamiltonian}
\end{equation}
	where
\begin{equation}
	S=\int d^{d+1}x\;\mathcal{L}.
\end{equation}

\vspace{6mm}

\subsection{Deformation of Canonical Commutation Relation}\mbox{}

	The new quantization, {\it noncommutative quantization}, which we will introduce bellow, can be considered as a deformation of the ordinary canonical quantization.
	This deformation is performed in terms of the star product extended to noncoinciding points\cite{Szabo:2001kg},
\begin{equation}
	f_1(x_1)\star f_2(x_2)\star \dots\star f_n(x_n)=\exp\left[\frac{i}{2}\theta^{ij}\Bigg({\displaystyle\sum^n_{a<b}}\partial_i^{x_a}\partial_j^{x_b}\Bigg)\right]f_1(x_1)f_2(x_2)\dots f_n(x_n)\label{star product}
\end{equation}
	for single variable functions.
	This product gives the ordinary Moyal star product when $x_1,\cdots,x_n\rightarrow x$.
	In this respect, we can consider this product as a natural generalization of the Moyal star product, which allows us to use this extended star product instead of the Moyal star product in the interaction term of eq.(\ref{lagrangian}).\newline
	In addition, we define the extended star product for functions with many variables as
\begin{equation}
\begin{array}{r}
	F(x_1,x_2,\cdots,x_m)\star G(y_1,y_2,\cdots,y_n)={\displaystyle \exp\left[\frac{i}{2}\theta^{ij}\Bigg(\sum_{a,b}\partial_i^{x_a}\partial_j^{y_b}\Bigg)\right]}\hspace{1.5cm}\\
	\times F(x_1,x_2,\cdots,x_m)G(y_1,y_2,\cdots,y_n).
\end{array}
\end{equation}

	Now we introduce a new quantization based on this extended star product: we promote $\phi$ and $\pi$ to operators by imposing the following {\it star-deformed canonical commutation relation},
\begin{equation}
\begin{array}{lcl}
	[\phi(t,\bx),\pi(t,\by)]_\star&=&\phi(t,\bx)\star\pi(t,\by)-\pi(t,\by)\star\phi(t,\bx)\\
	&=&i\delta^{(d)}(\bx-\by),\rule{0mm}{5mm}\\
	\strut[\phi(t,\bx),\phi(t,\by)]_\star&=&[\pi(t,\bx),\pi(t,\by)]_\star=0.\rule{0mm}{6mm}
\end{array}
	\label{nccr0}
\end{equation}
	This commutation relation reduces to the canonical commutation relation in the commutative limit $\theta^{ij}\rightarrow 0$.

	We shall see in the following that this quantization leads to a consistent QFT for scalar fields.

\vspace{9mm}

 \section{Noncommutative Quantization for Free Field}\mbox{}

	In this section, we consider the noncommutative quantization for free filed:
\begin{equation}
	S_0=\displaystyle \int d^{d+1}x\;\frac{1}{2}\Big[(\partial_\mu\phi)^2-m^2\phi^2\Big]\;.
\end{equation}
	This action gives
\begin{equation}
	(\Box+m^2)\phi=0.\label{eq of mot}
\end{equation}
	as equation of motion.
	From this equation of motion, we see that $\phi(x)$ can be expanded in the standard form:
\begin{equation}
	\phi(x)=\int \frac{d^dk}{(2\pi)^d}\frac{1}{\sqrt{2E_\bk}}\Big(a_\bk e^{-ikx}+a_\bk^\dagger e^{ikx}\Big)\Big\vert_{k^0=E_\bk=\sqrt{\bk^2+m^2}}.\label{fourier ex}
\end{equation}
	However, since $\phi(x)$ and $\pi(x)$ satisfy eq.(\ref{nccr0}) instead of the standard canonical commutation relation, $a_\bk$ and $a_\bk^\dagger$ do not obey the ordinary commutation relation for ladder operators.
	They satisfy rather
\begin{equation}
	\left\{
\begin{array}{l}
	e^{\frac{i}{2}k\theta p}a_\bk a_\bp^\dagger-e^{-\frac{i}{2}k\theta p}a_\bp^\dagger a_\bk=(2\pi)^d\delta^{(d)}(\bk-\bp),\\
	e^{-\frac{i}{2}k\theta p}a_\bk a_\bp-e^{\frac{i}{2}k\theta p}a_\bp a_\bk=0,\\
	e^{-\frac{i}{2}k\theta p}a_\bk^\dagger a_\bp^\dagger-e^{\frac{i}{2}k\theta p}a_\bp^\dagger a_\bk^\dagger=0,
\end{array}
	\right.\label{nccr1-1}
\end{equation}
	where
\begin{equation}
	k\theta p=k_i\theta^{ij}p_j=-p\theta k.
\end{equation}
	To make this expression easier, we introduce a star product in the momentum space as follows.
	Let $A_a$ and $\wp_a^{\mu}$ $(a=1,2,\dots,n)$ be $a_{\bp_a}$ and $p_a^\mu$, or $a_{\bp_a}^\dagger$ and $-p_a^\mu$ respectively.
	By use of this notation, we define a star product for $A_i$ as
\begin{equation}
	A_1\star A_2\star\cdots\star A_n=\exp\left[-\frac{i}{2}\Bigg({\displaystyle{\sum^n_{a<b}}}\wp_a\theta \wp_b\Bigg)\right]A_1A_2\cdots A_n.\label{star product-m}
\end{equation}
	It is easy to see that this product also satisfies associativity as well as the star product in position space (\ref{star product}).
	Useful examples are
\begin{equation}
\begin{array}{l}
	a_\bk\star a_\bp=e^{-\frac{i}{2}k\theta p}a_\bk a_\bp,\\
	a_\bk^\dagger\star a_\bp^\dagger=e^{-\frac{i}{2}k\theta p}a_\bk^\dagger a_\bp^\dagger=(a_\bp\star a_\bk)^\dagger,\rule{0mm}{5mm}\\
	a_\bk\star a_\bp^\dagger=e^{\frac{i}{2}k\theta p}a_\bk a_\bp^\dagger. \quad\rule{0mm}{5mm}
\end{array}
\end{equation}
	To clarify a relation between two star product eq.(\ref{star product}) and eq.(\ref{star product-m}), we notice the following relation is satisfied for the scalar field eq.(\ref{fourier ex}):
\begin{equation}
\begin{array}{rl}
	\phi(x)\star\phi(y)={\displaystyle \int\frac{d^dk d^dp}{(2\pi)^{2d}}\frac{1}{\sqrt{2E_\bk}}\frac{1}{\sqrt{2E_\bp}}\Big((a_\bk\star a_\bp)e^{-ikx-ipy}+(a_\bk\star a_\bp^\dagger)e^{-ikx+ipy}}\quad\\
	{\displaystyle +(a_\bk^\dagger\star a_\bp)e^{ikx-ipy}+(a_\bk^\dagger\star a_\bp^\dagger)e^{ikx+ipy}\Big)},
\end{array}
\end{equation}
	where the star product in LHS is given by eq.(\ref{star product}) and the star product in RHS is given by eq.(\ref{star product-m}).
	That is, the star product of functions is given by the star product of Fourier componets of them.
	From this equation, one can see eq.(\ref{star product-m}) is a proper definition of a star product in momentum space.

	By use of this star product in momentum space, eq.(\ref{nccr1-1}) can be rewritten in the next simple form:
\begin{equation}
\begin{array}{l}
	[a_\bk,a_\bp^\dagger]_\star=a_\bk\star a_\bp^\dagger-a_\bp^\dagger\star a_\bk=(2\pi)^d\delta^{(d)}(\bk-\bp),\\
	 \left[a_\bk,a_\bp\right]_\star=[a_\bk^\dagger,a_\bp^\dagger]_\star=0.\rule{0mm}{7mm}\label{nccr1-2}
\end{array}
\end{equation}
	These relation can be considered as a deformation of the commutation relation for the ordinary ladder operators by the star product (\ref{star product-m}).

\vspace{6mm}

 \subsection{Fock Space Representation}\mbox{}

	To get any physical informations from a quantum field theory, the quantum field must be represented on a Hilbert space.
	In the ordinary quantization of (commutative or noncommutative) free field theories, this is given by the Fock space, and their basis are well interpreted as energy-momentum eigenstates.
	Is there such a representation of the fields satisfying eq.(\ref{nccr0}) or equivalently eq.(\ref{nccr1-2})?
	This turns out to be the case.
	That is, we can construct a ``Fock space" on which the fields satisfying eq.(\ref{nccr0}) are represented\footnote{This ``Fock space" is the same one appeared in \cite{Balachandran:2005eb, Bu:2006ha}.} and we can interpret its basis as energy-momentum eigenstates.

	In the first place, we need a vacuum state $|0\rangle$:
\begin{equation}
	a_\bk|0\rangle=0,\quad \textrm{for all $a_\bk$.}
\end{equation}
	Then, $a_\bk^\dagger$ act on $|0\rangle$ to give the other basis vectors:
\begin{equation}
	|\bk_1,\bk_2,\cdots,\bk_n\rangle=a_{\bk_1}^\dagger\star a_{\bk_2}^\dagger\star\cdots\star a_{\bk_n}^\dagger|0\rangle.\label{Fock state}
\end{equation}
	These states satisfy Bose statistics from eq.(\ref{nccr1-2}).
	Any other state vectors are given by linear combinations of these vectors.
	The star product between these states is given by
\begin{equation}
\begin{array}{l}
	\langle \bp_1,\bp_2,\cdots,\bp_m|\star|\bk_1,\bk_2,\cdots,\bk_n\rangle\\
	=\langle0|a_{\bp_1}\star\cdots\star a_{\bp_m}\star a_{\bk_1}^\dagger\star\cdots\star a_{\bk_n}^\dagger|0\rangle=\delta_{m,n}\delta^{(d)}({\textstyle{\sum\bp_i-\sum\bk_i}}).\rule{0mm}{7mm}
\end{array}
	\label{inner product}
\end{equation}

	Next, we check that these basis vectors are energy-momentum eigenstates.
	The energy-momentum tensor is defined as an ordinary Noether's current in terms of a spacetime translation:
\begin{equation}
	T^{\mu\nu}=\partial^\mu\phi\partial^\nu\phi-g^{\mu\nu}\mathcal{L}_0.
\end{equation}
	From this, we can get the momentum operator:
\begin{equation}
\begin{array}{rcl}
	P^\mu&=&{\displaystyle{\int}}d^dx\;T^{0\mu}\\
	&=&{\displaystyle{\int}}d^dk\;\frac{\strut {\displaystyle{1}}}{\strut {\displaystyle{2}}}k^\mu\big(a_\bk^\dagger a_\bk+a_\bk a_\bk^\dagger\big)={\displaystyle{\int}}d^dk\;k^\mu a_\bk^\dagger a_\bk.
\end{array}
	\label{momentum operator}
\end{equation}
	To get last expression for $P^0$, we ignore an infinite constant.
	From this definition, it follows
\begin{equation}
\begin{array}{rcl}
	a_\bp^{(\dagger)}\star P^\mu&=&\displaystyle{\int d^d k\; k^\mu a_\bp^{(\dagger)}\star(a_\bk a_\bk^\dagger)}\\
	&=&\displaystyle{\int d^dk\; e^{-\frac{i}{2}(-)p\theta(k-k)}a_\bk a_\bk^\dagger}\\
	&=&a_\bp^{(\dagger)}P^\mu.\rule{0mm}{6mm}
\end{array}
\end{equation}
	For this $P^\mu$, we can easily verify
\begin{equation}
	P^\mu|0\rangle=0,\label{energy of grand state}
\end{equation}
\begin{equation}
	[P^\mu,a_\bk^\dagger]_\star=[P^\mu,a_\bk^\dagger]=k^\mu a_\bk^\dagger.\label{momentum commutation relation}
\end{equation}
	This result implies the state $|\bk_1,\bk_2,\cdots,\bk_n\rangle$ is the energy-momentum eigenstate whose eignevalue is given by $\sum_i k_i^\mu$.

\vspace{5mm}

\subsubsection{Star Product for State vectors}\mbox{}

	Here we comment on the definition of the star product for state vectors.	The star product between state vectors in eq.(\ref{inner product}) is understood by the definition of basis vectors eq.(\ref{Fock state}) and the star product for $a_\bk$ and $a_\bk^\dagger$ .
	We can also define the star product between states and other operators as follows.
	Defining the star product between $A_i$ in eq.(\ref{star product-m}) and an operator depending on space coordinates as
\begin{equation}
\begin{array}{l}
	\mathcal{O}(x_1,x_2,\cdots,x_n)\star A_1\star A_2\star \cdots \star A_m\\
	\hspace{1cm}=\exp\left[{\displaystyle \frac{1}{2}\Bigg(\sum_{i=1}^n\partial^{x_i}\Bigg)\theta\Bigg(\sum_{l=1}^m \wp_l\Bigg)}\right]\mathcal{O}(x_1,x_2,\cdots,x_n)(A_1\star A_2\star\cdots \star A_m),\rule{0mm}{1cm}
\end{array}\label{star product x-a}
\end{equation}
	the star product between $\mathcal{O}(x_1,\cdots,x_n)$ and basis vectors can be defined as
\begin{equation}
\begin{array}{l}
	\mathcal{O}(x_1,x_2,\cdots,x_n)\star|\bk_1,\bk_2,\cdots,\bk_m\rangle\\
	\hspace{1cm}=\mathcal{O}(x_1,x_2,\cdots,x_n)\star a_{\bk_1}^\dagger\star a_{\bk_2}^\dagger\star\cdots \star a_{\bk_m}^\dagger|0\rangle.\rule{0mm}{7mm}
\end{array}
	\label{star product x-o0}
\end{equation}
	The star product between an arbitrary state $|\alpha\rangle$ and an operator $\mathcal{O}(x_1,\cdots,x_n)$ is also well defined by expanding the state by basis vectors and applying eq.(\ref{star product x-o0}) to each basis vector.
	By use of the momentum operator (\ref{momentum operator}), this can be written as
\begin{equation}
	\mathcal{O}(x_1,x_2,\cdots,x_n)\star|\alpha\rangle=\mathcal{O}(x_1,x_2,\cdots,x_n)\exp\left[{\displaystyle -\frac{1}{2}}\Big({\textstyle \sum}\overleftarrow{\partial}^{x_i}\Big)\theta P\right]|\alpha\rangle.\label{star product x-o}
\end{equation}
	Here we adopt this equation as a definition of the star product between an operator $\mathcal{O}(x_1,\cdots,x_n)$ and an arbitrary state $|\alpha\rangle$ instead of (\ref{star product x-o0}).
	This definition is equivalent to (\ref{star product x-o0}) as long as one works in a free field theory.
	But (\ref{star product x-o0}) becomes somewhat ambiguous when one deals with an interacting field theory.
	On the other hand, (\ref{star product x-o}) can be always used as a definition as long as there is well defined momentum operator.
	Indeed, even if we include interaction term as eq.(\ref{lagrangian}), we can get well defined momentum operator\cite{Micu:2000xj,Gerhold:2000ik}.
	So we can define the star product between operator and arbitrary states for an interacting field theory by eq.(\ref{star product x-o}).
	In the same way, we can define the star product between $\langle\alpha|$ and $|\beta\rangle$ as
\begin{equation}
	\langle\alpha|\star|\beta\rangle=\langle\alpha|\exp\left(\overleftarrow{P}\theta\overrightarrow{P}\right)|\beta\rangle=\langle\alpha|\exp\left(P\theta P\right)|\beta\rangle.
\end{equation}
	Since $P\theta P=\theta^{ij}P^iP^j=0$, we find
\begin{equation}
	\langle\alpha|\star|\beta\rangle=\langle\alpha|\beta\rangle.\label{star product s-s}
\end{equation}

\vspace{6mm}

 \subsection{Propagator}\mbox{}
	
	Next, we calculate the Feynman propagator in our framework.
	All we have to do is to replace the ordinary product formally by the star product in each steps of the ordinary calculation.
	The result is
\begin{equation}
	\langle0|T_\star\phi(x)\phi(y)|0\rangle=\int\frac{d^{d+1}k}{(2\pi)^{d+1}}\frac{i}{k^2-m^2+i\epsilon}e^{-ik(x-y)}=:D_F(x-y),
\end{equation}
	where the {\it time-ordered star product} $T_\star$ is given by
\begin{equation}
	T_\star\phi(x)\phi(y)=\left\{
\begin{array}{l}
	\phi(x)\star\phi(y),\textrm{ when $x^0>y^0$,}\\
	\phi(y)\star\phi(x),\textrm{ when $y^0>x^0$.}
\end{array}
	\right.\label{time-ordered}
\end{equation}

	Here we introduce the {\it normal-ordered star product} $N_\star$ for later use.
	For $a_\bk^{(\dagger)}$, this is given by, for example,
\begin{equation}
	N_\star a_\bk a_\bp^\dagger a_\bq=a_\bp^\dagger\star a_\bk\star a_\bq.
\end{equation}
	For $\phi(x)$, this reads
\begin{equation}
\begin{array}{rcl}
	N_\star\phi(x)\phi(y)&=&\phi^{(+)}(x)\star\phi^{(+)}(y)+\phi^{(-)}(x)\star\phi^{(+)}(y)\\
	&&+\phi^{(-)}(y)\star\phi^{(+)}(x)+\phi^{(-)}(x)\star\phi^{(-)}(y),\rule{0mm}{5mm}
\end{array}
\end{equation}
	where
\begin{equation}
	\left\{
\begin{array}{l}
	\phi^{(+)}(x)={\displaystyle{\int}}\frac{\displaystyle{d^dk}}{\displaystyle{(2\pi)^d}}\frac{\displaystyle{1}}{\displaystyle{\sqrt{2E_\bk}}}a_\bk e^{-ikx}\\
	\phi^{(-)}(x)={\displaystyle{\int}}\frac{\displaystyle{d^dk}}{\displaystyle{(2\pi)^d}}\frac{\displaystyle{1}}{\displaystyle{\sqrt{2E_\bk}}}a_\bk^\dagger e^{ikx}.
\end{array}
	\right.
\end{equation}

\vspace{9mm}

\section{Noncommutative Quantization for Interacting Field}\mbox{}

	In this section, we consider the noncommutative quantization of an interacting field theory eq.(\ref{lagrangian}).
	The action of this theory gives the following equation of motion:
\begin{equation}
	(\Box+m^2)\phi-\sum_{n=3}\frac{\lambda_n}{(n-1)!}\overbrace{\phi\star\phi\star\cdots\star\phi}^{n-1}=0.\label{eq of mot 1}
\end{equation}
	By use of canonical momentum $\pi$ in eq.(\ref{canonical momentum}), which satisfies the star-deformed commutation relation (\ref{nccr0}), and Hamiltonian (\ref{hamiltonian}), this equation of motion can be rewritten as
\begin{equation}
\begin{array}{l}
	\dot{\phi}(x)=i[H,\phi(x)]_\star=i[H,\phi(x)],\\
	\dot{\pi}(x)=i[H,\pi(x)]_\star=i[H,\pi(x)].\rule{0mm}{5mm}
\end{array}
	\label{eq of mot 2}
\end{equation}
	The second equaities for each equations hold because the Hamiltonian $H$ does not depend on coordinate so that
\begin{equation}
\begin{array}{ll}
	H\star\phi(x)=H\phi(x),&\phi(x)\star H=\phi(x)H,\\
	\strut H\star\pi(x)=H\pi(x),&\pi(x)\star H=\pi(x)H.\rule{0mm}{5mm}
\end{array}
\end{equation}
	From eq.(\ref{eq of mot 2}), we can see that the way of time evolution of our theory is exactly the same as ordinary quantum field theories.
	In particular, this implies that we can use the interaction picture to describe the interacting field theory, that is, one can perform a perturbation expansion to calculate Green's functions or S-matrix elements when the interaction term $\mathcal{L}_\textrm{\scriptsize{int}}$ can be treated as perturbation.

\vspace{6mm}

 \subsection{Green's Function}\mbox{}

	We firstly consider Green's function.
	In the interaction picture, the n-point Green's function appropriate for our quantization scheme is given by
\begin{equation}
	G_\star^{(n)}(x_1,x_2,\cdots,x_n)=\frac{\strut Z^{-\frac{n}{2}}\langle 0|T_\star\phi_I(x_1)\phi_I(x_2)\cdots\phi_I(x_n)\exp[-i\int dt\;H_\textrm{\scriptsize{int}}(t)]|0\rangle}{\strut\langle 0|T_\star\exp[-i\int dt\;H_\textrm{\scriptsize{int}}(t)]|0\rangle}, \label{green}
\end{equation}
	where $H_\textrm{\scriptsize{int}}(t)$ is an interaction part of a Hamiltonian,
\begin{equation}
	H_\textrm{\scriptsize{int}}(t)=\sum_{n=3}^\infty\frac{\lambda_n}{n!}\int d^dx\overbrace{\phi_I(x)\star\phi_I(x)\cdots\star\phi_I(x)}^n,
\end{equation}
	and $Z$ is a field-strength renormalization\footnote{The value of $Z$ is determined in the same way as an ordinary QFT. That is, it is fixed as the residue of the 2-point function at a pole to be 1. From the result of this section, we see that it can be determined definitely when the corresponding commutative theory is a renormalizable theory.}.
	The subscript $I$ of $\phi_I(x)$ means $\phi_I(x)$ is the interaction picture field and it is expanded in the same form as the free field eq.(\ref{fourier ex}) by use of  $a_\bk$ and $a_\bk^\dagger$ which satisfy eq.(\ref{nccr1-2}).
	To calculate this Green's function perturbatively, we need Wick's theorem:
\begin{equation}
\begin{array}{l}
	T_\star\{\phi_I(x_1)\phi_I(x_2)\cdots\phi_I(x_n)\}\\
	\quad=N_\star\{\phi_I(x_1)\phi_I(x_2)\cdots\phi_I(x_n)+\wick{1}{<1{\phi_I}(x_1)>1{\phi_I}(x_2)}\cdots\phi_I(x_n)\\
	\hspace{5cm}+\textrm{ all other possible contractions}\},\rule{0mm}{5mm}
\end{array}
	\label{wick}
\end{equation}
	where the contraction is defined as
\begin{equation}
	\wick{2}{<1{\phi_I}(x)>1{\phi_I}(y)}=D_F(x-y).
\end{equation}
	and, for example,
\begin{equation}
\begin{array}{lcl}
	N_\star\{\wick{1}{<1{\phi_I}(x_1)>1{\phi_I}(x_2)}\phi_I(x_3)\phi_I(x_4)\}&\textrm{means}&D_F(x_1-x_2)\cdot N_\star\{\phi_I(x_3)\phi_I(x_4)\},\\
	N_\star\{\wick{1}{<1{\phi_I}(x_1)\phi_I(x_2)>1{\phi_I}(x_3)}\phi_I(x_4)\}&\textrm{means}&D_F(x_1-x_3)\cdot N_\star\{\phi_I(x_2)\phi_I(x_4)\}\rule{0mm}{10mm}
\end{array}
\end{equation}
	To prove this Wick's theorem, one needs to notice the following relation:
\begin{equation}
\begin{array}{l}
	\mathcal{O}(x_1,\cdots,x_n)\star D_F(y_1-y_2)\star\mathcal{O}'(z_1,\cdots,z_m)\\
	=D_F(y_1-y_2)\cdot\Big(\mathcal{O}(x_1,\cdots,x_n)\star\mathcal{O}'(z_1,\cdots,z_m)\Big),\rule{0mm}{7mm}
\end{array}
	\label{star product for propagator}
\end{equation}
	where $\mathcal{O}(x_1,\cdots,x_n)$ and $\mathcal{O}'(z_1,\cdots,z_m)$ are arbitrary operators which depend on coordinates.
	This relation results immediately from
\begin{equation}
	(\partial^{y_1}+\partial^{y_2})D_F(y_1-y_2)=0.
\end{equation}
	Except for this relation (\ref{star product for propagator}), the proof of eq.(\ref{wick}) proceeds almost in the same way as the standard case.
	Only we have to do is to replace all the products in the proof of the ordinary Wick's theorem by the star product eq.(\ref{star product}) formally.

	As an examle of a calculation of the Green's function, let us consider the noncommutative $\phi^4$ theory:
\begin{equation}
	H_\textrm{\scriptsize{int}}=\frac{\lambda_4}{4!}\int d^dx\;\phi_I(x)\star\phi_I(x)\star\phi_I(x)\star\phi_I(x).
\end{equation}
	For 2-point function of this theory, we must calculate
\begin{equation}
	\langle 0|\;T_\star\phi_I(x_1)\phi_I(x_2)\exp\Big[-i\int dt\;H_\textrm{\scriptsize{int}}(t)\Big]|0\rangle.\label{2 point}
\end{equation}
	A typical term to be calculated is evaluated by use of the Wick's theorem (\ref{wick}) as follows:
\begin{equation}
\begin{array}{l}
	-i\frac{\displaystyle{\lambda_4}}{\displaystyle{4!}}{\displaystyle{\int}}d^{d+1}x\;\langle 0|\;T_\star\{\phi_I(x_1)\phi_I(x_2)\phi_I(x)\phi_I(x)\phi_I(x)\phi_I(x)\}|0\rangle \\
	=-i\frac{\displaystyle{\lambda_4}}{\displaystyle{4!}}\cdot3\cdot D_F(x_1-x_2){\displaystyle{\int}}d^{d+1}x\;D_F(x-x)D_F(x-x)\\
	\hspace{1cm}-i\frac{\displaystyle{\lambda_4}}{\displaystyle{4!}}\cdot12\cdot{\displaystyle{\int}}d^{d+1}x\;D_F(x_1-x)D_F(x_2-x)D_F(x-x).
\end{array}
\end{equation}
	Notice that this result is exactly the same as the corresponding one of commutative $\phi^4$ theory.
	This is because, different from the standard noncommutative QFT, $\phi_I(x)$ are mutually commutative because of eq.(\ref{nccr0}) in the interaction term.
	Since this property holds independently of an order of perturbation, we can see that the Green's function eq.(\ref{green}) has the same value as the corresponding Green's function of commutative $\phi^4$ theory in all orders of perturbation.
	It is easy to see that this property is also independent of the form of the interaction term as long as fields in it are multiplied by the star product.\newline
	This equality between two Green's functions in any orders of perturbation implies, in particular, that the renormalization scheme can be used as well for our theory, provided the corresponding commutative theory is a renormalizable theory.

\vspace{6mm}

 \subsection{S-Matrix}\mbox{}

	Next, we consider S-matrix elements.
	To define a S-matrix, we need notions of asymptotic fields and the asymptotic completenss.
	We can introduce these notions in the same way as the ordinary QFT.
	That is, asymptotic fields are introduced in the Heisenberg picture as follows.
\begin{equation}
\begin{CD}
	\phi(x)@>\textrm{\scriptsize w}>x^0\rightarrow-\infty>\sqrt{Z}\phi^\textrm{\scriptsize{in}}(x)=\sqrt{Z}{\displaystyle{\int \frac{d^{d}k}{(2\pi)^d}\frac{1}{\sqrt{2E_\bk}}}}\Big[a^\textrm{\scriptsize{in}}_\bk e^{-ikx}+a_\bk^{\textrm{\scriptsize{in}}\dagger} e^{ikx}\Big]\\
	@>\textrm{\scriptsize w}>x^0\rightarrow+\infty>\sqrt{Z}\phi^\textrm{\scriptsize{out}}(x)=\sqrt{Z}{\displaystyle{\int \frac{d^dk}{(2\pi)^d}\frac{1}{\sqrt{2E_\bk}}}}\Big[a^\textrm{\scriptsize{out}}_\bk e^{-ikx}+a_\bk^{\textrm{\scriptsize{out}}\dagger} e^{ikx}\Big],
\end{CD}
\end{equation}
	where w over the arrows means it is a weak limit.
	These asymptotic fields satisfy the next equations:
\begin{equation}
	(\Box+\overline{m}^2)\phi^\textrm{\scriptsize{as}}(x)=0,\quad\textrm{(as$=$in or out, $\overline{m}$ : renormalized mass)}
\end{equation}
	and
\begin{equation}
\begin{array}{l}
	[a^\textrm{\scriptsize{as}}_\bk,a_\bp^{\textrm{\scriptsize{as}}\dagger}]_\star=(2\pi)^d\delta^{(d)}(\bk-\bp),\\
	 \left[a_\bk^\textrm{\scriptsize{as}},a_\bp^\textrm{\scriptsize{as}}\right]_\star=[a_\bk^{\textrm{\scriptsize{as}}\dagger},a_\bp^{\textrm{\scriptsize{as}}\dagger}]_\star=0.\rule{0mm}{7mm}\label{anccr}
\end{array}
\end{equation}
	Denoting the Fock space constructed by $a_\bk^\textrm{\scriptsize{as}}$ as $\mathcal{V}^\textrm{\scriptsize{as}}$ and the Hilbert space on which $\phi(x)$ is represented as $\mathcal{V}$, the asymptotic completeness is given by
\begin{equation}
	\mathcal{V}^\textrm{\scriptsize{in}}=\mathcal{V}^\textrm{\scriptsize{out}}=\mathcal{V}.
\end{equation}
	In these settings, S-matrix elements are defined as
\begin{equation}
	\langle\alpha\;\textrm{out}|\star|\beta\;\textrm{in}\rangle=S_{\alpha\beta},
\end{equation}
	where $|\alpha\;\textrm{as}\rangle\in\mathcal{V}^\textrm{\scriptsize{as}}$.

	In the ordinary QFT, to evaluate a S-matrix element from the corresponding Green's function, one uses LSZ reduction formula.
	For our theory, this formula holds if we replace the Green's function in it by that of our theory eq.(\ref{green}).
\begin{equation}
\begin{array}{l}
	\langle\bk_1,\bk_2,\cdots,\bk_m\;\textrm{out}|\star|\bp_1,\bp_2,\cdots,\bp_n\;\textrm{in}\rangle\\
	=\displaystyle{\prod^m_{a=1}}\Big(i\int d^{d+1}x_a\;f^{\ast}_{\bk_a}(x_a)(\Box_{x_a}+\overline{m}^2)\Big)\prod^n_{b=1}\Big(i\int d^{d+1}y_b\;f_{\bp_b}(x_b)(\Box_{y_b}+\overline{m}^2)\Big)\\
	\qquad\times G^{(m+n)}_\star(x_1,x_2,\cdots,x_m,y_1,y_2,\cdots,y_n),\rule{0mm}{5mm}
\end{array}
	\label{LSZ}
\end{equation}
	where $\bk_a\ne\bp_b$ for any $a,b$ and
\begin{equation}
	f_\bk(x)=\frac{e^{-ikx}}{\sqrt{2k_0}}\;\Big\vert_{k^0=\sqrt{\bk^2+\overline{m}^2}}.
\end{equation}
	Notice that products between $f_{\bk_a}^{(\ast)}$ and $G_\star^{(m+n)}$ are not star products but ordinary products\footnote{In fact, even if we insert the star product between $f_{\bk_i}^{(\ast)}$ and $G_\star^{(m+n)}$ in the RHS of (\ref{LSZ}), it disappears to give the same expression as the RHS of (\ref{LSZ}). This is because the difference between them is only total derivative terms and it is integrated out from the expression.}.
	The proof of this formula is almost the same as the commutative case again.
	In section 4.1, we see that Green's functions of our theory have exactly the same value as those of the ordinary commutative theory.
	Eq.(\ref{LSZ}) suggests that this is also the case for S-matrix elements, that is, there is no difference between the value of S-matrix elements of our theory and those of the ordinary commutative theory.

\vspace{9mm}

\section{Noncommutativity Cancellation}\mbox{}

	As we have seen in the previous section, Green's functions and S-matirix elements calculated perturbatively in our theory are exactly the same as those of a commuatative theory, even though the action includes a noncommutative interaction term.
	This is an outstanding property of the noncommutative quantization.
	So we investigate this structure in detail in this section.

\vspace{6mm}

 \subsection{Cancellation in Perturbation Calculation}\mbox{}

	The calculation in the previous section suggests that the noncommuative parameter appearing in the quantization procedure and the noncommuative parameter in an interaction term cancel out each other and does not appear in Green's functions or S-matrix elements.
	To see how this cancellation happens, we shall use different noncommuative parameters for a quantization and an interaction term in this section.
	To this end, we denote a star product in terms of a noncommutative parameter $\theta^{ij}$ as $\star_\theta$ and a field quantized by a star commutator $[\;,\;]_{\star_\theta}$ as $\phi^\theta(x)$.
	As an example, let us consider $\phi^3$ theory which uses a $\theta$-star product $\star_\theta$ for the quantization and a $\tilde\theta$-star product $\star_{\tilde\theta}$ for the interaction:
\begin{equation}
\begin{array}{rcl}
	H_\textrm{\scriptsize{int}}^{(\theta,\tilde\theta)}&=&{\displaystyle \frac{\lambda_3}{3!}\int d^dx\;\phi^\theta_I(x)\star_{\tilde\theta}\phi^\theta_I(x)\star_{\tilde\theta}\phi^\theta_I(x)}\\
	&=&{\displaystyle \frac{\lambda_3}{3!}\int d^dx\;\exp\left[{\displaystyle \frac{i}{2}\sum_{a<b}}\partial^a\tilde\theta\partial^b\right]}{\displaystyle \phi^\theta_I(x_1)\phi^\theta_I(x_2)\phi^\theta_I(x_3)\Big\vert_{x_1,x_2,x_3\rightarrow x}},
\end{array}
	\label{general interaction}
\end{equation}
	where $\phi^\theta_I(x)$ satisfies
\begin{equation}
	[\phi^\theta_I(t,\bx),\phi^\theta_I(t,\by)]_{\star_\theta}=\phi^\theta_I(x)\star_\theta\phi^\theta_I(y)-\phi^\theta_I(y)\star_\theta\phi^\theta_I(x)=0.\label{noncommutative commutation}
\end{equation}
	Let us call this theory as $(\theta,\tilde\theta)$-theory.

	In this notation, the ordinary commutative $\phi^3$ theory is given by the case $\theta=\tilde\theta=0$:
\begin{equation}
\begin{array}{rcl}
	H_\textrm{\scriptsize{int}}^{(0,0)}&=&{\displaystyle \frac{\lambda_3}{3!}\int d^dx\;\phi^0_I(x)\star_{0}\phi^0_I(x)\star_{0}\phi^0_I(x)}\\
	&=&{\displaystyle \frac{\strut\lambda_3}{3!}\int d^dx\;\phi^0_I(x)\phi^0_I(x)\phi^0_I(x)},\rule{0mm}{6mm}
\end{array}
	\label{commutative interaction}
\end{equation}
\begin{equation}
	[\phi^0_I(t,\bx),\phi^0_I(t,\by)]=0.\label{commutative commutation}
\end{equation}
	These indicate that three $\phi^0_I(x)$ are commutative each other in the interaction term so that whichever $\phi^0_I(x)$ we use as a partner of a contraction, its result does not change.

	Next, the ordinary noncommutative $\phi^3$ theory using the canonical quantization is given by the case $\theta=0,\tilde\theta\ne0$:
\begin{equation}
\begin{array}{rcl}
	H_\textrm{\scriptsize{int}}^{(0,\tilde\theta)}&=&{\displaystyle \frac{\lambda_3}{3!}\int d^dx\;\phi^0_I(x)\star_{\tilde\theta}\phi^0_I(x)\star_{\tilde\theta}\phi^0_I(x)}\\
	&=&{\displaystyle \frac{\lambda_3}{3!}\int d^dx\;\exp\left[{\displaystyle \frac{i}{2}\sum_{a<b}}\partial^a\tilde\theta\partial^b\right]}{\displaystyle \phi^0_I(x_1)\phi^0_I(x_2)\phi^0_I(x_3)\Big\vert_{x_1,x_2,x_3\rightarrow x}},
\end{array}
	\label{noncommutative interaction}
\end{equation}
	where $\phi^0_I(x)$ satisfies eq.(\ref{commutative commutation}) again.
	In this case, $\phi^0_I(x)$ cannot commute in the interaction term because of the exponential operator.
	This indicates that when taking a contraction in a perturbation calculation, one must care about which $\phi^0_I(x)$ ($\phi^0_I(x_1)$, $\phi^0_I(x_2)$ or $\phi^0_I(x_3)$) is contracted with another operator.
	This property leads to the distinction between planer and nonplaner diagram, and gives nontrivial noncommutative perturbative dynamics such as UV/IR mixing\cite{Minwalla:1999px, VanRaamsdonk:2000rr, Aref'eva:1999sn}.

	For a general $(\theta,\tilde\theta)$ case, to adapt to the commutation relation (\ref{noncommutative commutation}) we rewrite the intraction term (\ref{general interaction}) as follows:
\begin{equation}
	H_\textrm{\scriptsize{int}}^{(\theta,\tilde\theta)}={\displaystyle \frac{\lambda_3}{3!}\int d^dx\;\exp\left[{\displaystyle \frac{i}{2}}\sum_{a<b}\partial^a\Theta\partial^b\right]}{\displaystyle \phi^\theta_I(x_1)\star_\theta\phi^\theta_I(x_2)\star_\theta\phi^\theta_I(x_3)\Big\vert_{x_1,x_2,x_3\rightarrow x}},\label{general interaction2}
\end{equation}
	where $\Theta^{ij}=\tilde\theta^{ij}-\theta^{ij}$.
	From this expression and the commutation relation (\ref{noncommutative commutation}), we can see that the exponential operator which depends on $\Theta^{ij}$ prevents $\phi^\theta_I(x_1)$, $\phi^\theta_I(x_2)$ and $\phi^\theta_I(x_3)$ from commuting each other in the interaction term just the same way as the ordinary noncommutative case (\ref{noncommutative interaction}).
	The only exception is the case $\Theta^{ij}=0$, namely the case of the theory we have considered, in which the exponential operator vanishes.
	In this case, three $\phi^\theta_I(x)$ can commute freely in the interaction term.
	In addition, it is easy to see that, aside from properties concerning the space-time symmetry, the difference between the term $\phi^0_I(x)\phi^0_I(x)\phi^0_I(x)$ and the term $\phi^\theta_I(x)\star_\theta\phi^\theta_I(x)\star_\theta\phi^\theta_I(x)$ is only appearance as long as they are equipped with proper Fock spaces, that is, the ordinary Fock space for the former and $\theta$-Fock space in section 3.1 for the lattar.
	We can always move from one to the other by replacing all products formally.
	Then, it follows that, in the case $\Theta^{ij}=0$, the result of a perturbation calculation does not depend on which $\phi^\theta_I(x)$ in the interaction term one chooses as a contraction partner.
	As a result, all the $(\theta,\theta)$-theories, i.e., the theories with $\Theta^{ij}=0$, give the same Green's functions and S-matrix elements as those of the ordinary commutative theory.
	These are the reason that the noncommutative parameter does not appear explicitly in the perturbation calculation in section 4.
	It is clear that this disucussion and conclusion hold for other scalar field theories such as $\phi^4$ theory.
	
	From the discussion above, it is almost evident that a theory which has nonzero $\Theta^{ij}$ results in a theory equivalent to the ordinary noncommutative theory which has a noncommutative parameter $\Theta^{ij}$, i.e., $(0,\Theta)$-theory.
	In the case where $\Theta^{ij}$ is nonzero, the exponential operator in eq.(\ref{general interaction2}) plays exactly the same role as the exponential operator in eq.(\ref{noncommutative interaction}) in a perturbation calculation.
	This structure suggests that it is possible to classify general $(\theta,\tilde\theta)$-theories in terms of $\Theta^{ij}$.
	The theories which have the same $\Theta^{ij}$ give the same Green's functions and S-matrix elements.
	This means that, in particular, $(0,\theta)$-theory, i.e., the theory which has a noncommutative interaction term and is quantized by the standard canonical quantization procedure, is equivalent to $(-\theta,0)$ theory, i.e., the theory which has a commutative interaction term and is quantized by the $-\theta$-star product $\star_{-\theta}$.
	In other words, by using the noncommutative quantization, the noncommutativity can be incorporated in the process of quantization, rather than in the action.

\vspace{6mm}

 \subsection{Nonperturbative Correspondence}\mbox{}

	We have seen the noncommutativity cancellation between the interaction term and the quantization procedure in a perturbation calculation.
	In fact, this property persists even nonperturbatively.
	We can prove this by establishing a map between the ordinary commutative QFT and our new QFT.
	That is, we can establish a map from a commutative QFT,
\begin{equation}
\begin{array}{l}
	\displaystyle{\mathcal{L}^0=\frac{1}{2}\Big[(\partial_\mu\phi^0)^2-m^2(\phi^0)^2\Big]+\sum_{n=3}\frac{\lambda_n}{n!}(\phi^0)^n}\\
	\displaystyle{\longrightarrow\textrm{Eq. of Mot. : }(\Box+m^2)\phi^0-\sum_{n=3}\frac{\lambda_n}{(n-1)!}(\phi^0)^{n-1}}=0,
\end{array}
	\label{old theory}
\end{equation}
	equipped with the standard canonical commutation relation (i.e., $(0,0)$-theory in section 5.1), to our new QFT,
\begin{equation}
\begin{array}{l}
	\displaystyle{\mathcal{L}^\theta=\frac{1}{2}\Big[(\partial_\mu\phi^\theta)^2-m^2(\phi^\theta)^2\Big]+\sum_{n=3}\frac{\lambda_n}{n!}\overbrace{\phi^\theta\star\cdots\star\phi^\theta}^n}\\
	\displaystyle{\longrightarrow\textrm{Eq. of Mot. : }(\Box+m^2)\phi^\theta-\sum_{n=3}\frac{\lambda_n}{(n-1)!}\overbrace{\phi^\theta\star\cdots\star\phi^\theta}^{n-1}}=0,
\end{array}
	\label{new theory}
\end{equation}
	equipped with the star-deformed commutation relation (\ref{nccr0}) (i.e., $(\theta, \theta)$-theory in section 5.1).
	In particular, we can find out a map from $\phi^0(x)$ to $\phi^\theta(x)$.
	To write down this map, we use the ordinary momentum operator of $(0,0)$-theory,
\begin{equation}
	\mathcal{P}^{\mu}=\int d^d x\;\Big[\partial^0\phi^0\partial^\mu\phi^0-g^{0\mu}\mathcal{L}^0\Big].\label{old momentum}
\end{equation}

	By use of this momentum operator, the map from $\phi^0$ to $\phi^{\theta}$ is given by
\begin{equation}
	\phi^\theta(x)=\exp\left(-\frac{1}{2}\mathcal{P}\theta\partial\right)\phi^0(x).\label{theory map}
\end{equation}
	It is easy to see that this $\phi^\theta$ and the corresponding canonical momentum $\pi^\theta$ satisfy the star-deformed commutation relations (\ref{nccr0}), provided that $\phi^0$ and the corresponding canonical momentum $\pi^0$ satisfy the standard canonical commutation relation.
	For example, from
\begin{equation}
\begin{array}{l}
	\phi^\theta(x)\star\phi^\theta(y)\\
	=\displaystyle{\exp\left(\frac{i}{2}\partial^x\theta\partial^y\right)\left[\exp\left(-\frac{1}{2}\mathcal{P}\theta\partial^x\right)\phi^0(x)\right]\left[\exp\left(-\frac{1}{2}\mathcal{P}\theta\partial^y\right)\phi^0(y)\right]}\rule{0mm}{10mm}\\
	\displaystyle{=\exp\left(\frac{i}{2}\partial^x\theta\partial^y\right)\exp\left(-\frac{1}{2}\mathcal{P}\theta\partial^x\right)\exp\left(-\frac{1}{2}(\mathcal{P}+i\partial^x)\theta\partial^y\right)\phi^0(x)\phi^0(y)}\rule{0mm}{10mm}\\
	\displaystyle{=\exp\left(-\frac{1}{2}\mathcal{P}\theta(\partial^x+\partial^y)\right)\phi^0(x)\phi^0(y)},\rule{0mm}{10mm}
\end{array}
	\label{map of products}
\end{equation}
	it follows
\begin{equation}
	[\phi^\theta(x),\phi^\theta(y)]_\star=\exp\left(-\frac{1}{2}\mathcal{P}\theta(\partial^x+\partial^y)\right)[\phi^0(x),\phi^0(y)].
\end{equation}
	Here we use a standard formula $e^{a\mathcal{P}}\phi^0(x)e^{-a\mathcal{P}}=e^{-ia\partial}\phi^0(x)$ to derive eq.(\ref{map of products}).
	We can easily verify the other two relations in the same way,
\begin{equation}
\begin{array}{l}
	{\displaystyle [\pi^\theta(x),\pi^\theta(y)]_\star=\exp\left(-\frac{1}{2}\mathcal{P}\theta(\partial^x+\partial^y)\right)[\pi^0(x),\pi^0(y)]}\\
	{\displaystyle [\phi^\theta(x),\pi^\theta(y)]_\star=\exp\left(-\frac{1}{2}\mathcal{P}\theta(\partial^x+\partial^y)\right)[\phi^0(x),\pi^0(y)].}\rule{0mm}{10mm}
\end{array}
\end{equation}
	These relations indicate that, if $\phi^0$ and $\pi^0$ satisfy the canonical commutation relation, then $\phi^\theta$ and $\pi^\theta$ satisfy the star-deformed canonical commutation relation.

	The equation of motion in eq.(\ref{new theory}) is also derived from the equation of motion in eq.(\ref{old theory}) by the map (\ref{theory map}).
\begin{equation}
\begin{array}{l}
	\displaystyle{(\Box+m^2)\phi^\theta(x)-\sum_{n=3}\frac{\lambda_n}{(n-1)!}\overbrace{\phi^\theta(x)\star\cdots\star\phi^\theta(x)}^{n-1}}\\
	\displaystyle{=(\Box+m^2)\exp\left(-\frac{1}{2}\mathcal{P}\theta\partial\right)\phi^0(x)}\rule{0mm}{8mm}\\
	\hspace{1cm}\displaystyle{-\sum_{n=3}\frac{\lambda_n}{(n-1)!}\exp\left[-\frac{1}{2}\mathcal{P}\theta\Big(\sum_{a=1}^{n-1}\partial^{x_a}\Big)\right]\phi^0(x_1)\cdots\phi^0(x_{n-1})\Big\vert_{x_a\rightarrow x}}\rule{0mm}{8mm}\\
	\displaystyle{=(\Box+m^2)\exp\left(-\frac{1}{2}\mathcal{P}\theta\partial\right)\phi^0(x)-\sum_{n=3}\frac{\lambda_n}{(n-1)!}\exp\left(-\frac{1}{2}\mathcal{P}\theta\partial\right)(\phi^0(x))^{n-1}}\rule{0mm}{8mm}\\
	\displaystyle{=\exp\left(-\frac{1}{2}\mathcal{P}\theta\partial\right)\left[(\Box+m^2)\phi^0(x)-\sum_{n=3}\frac{\lambda_n}{(n-1)!}(\phi^0(x))^{n-1}\right],}\rule{0mm}{10mm}
\end{array}
	\label{map of eq. of mot.}
\end{equation}
	where we use a relation
\begin{equation}
	(\partial^{x_1}+\cdots+\partial^{x_n})f^1(x_1)\cdots f^n(x_n)\Big\vert_{x_a\rightarrow x}=\partial^x(f^1(x)\cdots f^n(x)),\label{relation}
\end{equation}
	to get the third line.
	Eq.(\ref{map of eq. of mot.}) implies that, if the eqation of motion in eq.(\ref{old theory}) is satisfied, then the equation of motion in eq.(\ref{new theory}) is also satisfied.
	These results indicate that the map (\ref{theory map}) exactly gives a map from $\phi^0$ in eq.(\ref{old theory}) to $\phi^\theta$ in eq.(\ref{new theory}).

	The map (\ref{theory map}) implies that a Hilbert space $\mathcal{V}_0$ on which $\phi^0$ is represented can be also used to represent $\phi^\theta$.
	Moreover, properties of a state in $\mathcal{V}_0$ are not changed by this map.
	To clarify the meaning of this statement, we first notice that the momentum operator of $(\theta, \theta)$-theory,
\begin{equation}
	\mathcal{P}_\theta^\mu=\int d^dx\;\Big[\partial^0\phi^\theta\partial^\mu\phi^\theta-g^{0\mu}\mathcal{L}^\theta\Big],\label{theta momentum}
\end{equation}
	is equal to $\mathcal{P}^\mu$.
	This momentum operator is defined as a Noether current in terms of a spacetime translation for the Lagrangian in eq.(\ref{new theory}).
	Indeed, using the map (\ref{theory map}) and the relation (\ref{relation}), RHS of this definition is rewritten as
\begin{equation}
	\int d^dx\;\exp\Big(-\frac{1}{2}\mathcal{P}\theta\partial\Big)\Big[\partial^0\phi^0\partial^\mu\phi^0-g^{0\mu}\mathcal{L}^0\Big]
\end{equation}
	and, ignoring total derivative terms in the integrand, this is equal to $\mathcal{P}^\mu$.
	Based on this equality, one can see that the following two actions of operators to a state $|\alpha\rangle\in\mathcal{V}_0$ are equivalent.
\begin{equation}
\begin{array}{c}
\begin{array}{l}
	\textrm{Ordinary action of an operator $\mathcal{O}[\phi^0]$\hspace{3mm}:}\hspace{5mm}\mathcal{O}[\phi^0]|\alpha\rangle\\
	\textrm{Star-action of an operator $\mathcal{O}_\star[\phi^\theta]$\hspace{3mm}:}\hspace{5mm}\mathcal{O}_\star[\phi^\theta]\star|\alpha\rangle,\rule{0mm}{7mm}
\end{array}\\
	\Rightarrow\quad \mathcal{O}[\phi^0]|\alpha\rangle=\mathcal{O}_\star[\phi^\theta]\star|\alpha\rangle,\rule{0mm}{8mm}
\end{array}
	\label{two actions}
\end{equation}
	where $\mathcal{O}[\phi^0]$ is arbitrary operator constructed from $\phi^0$ by the ordinary product, $\mathcal{O}_\star[\phi^\theta]$ is an operator replacing all $\phi^0$ and all their products in $\mathcal{O}[\phi^0]$ by $\phi^\theta$ and the star product, and a star-action means the action defined in eq.(\ref{star product x-o}).
	Notice that $P^\mu$ in eq.(\ref{star product x-o}) is given by $\mathcal{P}_\theta^\mu$ here.

	For example, let us consider the case $\mathcal{O}[\phi^0]=\phi^0(x_1)\phi^0(x_2)\cdots\phi^0(x_n)$.
	The corresponding operator $\mathcal{O}_\star[\phi^\theta]$ is $\phi^\theta(x_1)\star\phi^\theta(x_2)\star\cdots\star\phi^\theta(x_n)$, and by eq.(\ref{theory map}) this is given in terms of $\phi^0$ by
\begin{equation}
	\phi^\theta(x_1)\star\phi^\theta(x_2)\star\cdots\star\phi^\theta(x_n)=\phi^0(x_1)\phi^0(x_2)\cdots\phi^0(x_n)\exp\left[-\frac{1}{2}\mathcal{P}\theta\Big(\sum_{i=1}^n\overleftarrow{\partial}^{x_i}\Big)\right].
\end{equation}
	Using $\mathcal{P}^\mu=\mathcal{P}_\theta^\mu$, one can easily find the star-action of this operator to $|\alpha\rangle$ is equal to the ordinary action of $\phi^0(x_1)\phi^0(x_2)\cdots\phi^0(x_n)$,
\begin{equation}
	\phi^\theta(x_1)\star\phi^\theta(x_2)\star\cdots\star\phi^\theta(x_n)\star|\alpha\rangle=\phi^0(x_1)\phi^0(x_2)\cdots\phi^0(x_n)|\alpha\rangle.
\end{equation}
	This equality between two actions in eq.(\ref{two actions}) means that $|\alpha\rangle$ carries the same quantum numbers both in $(0,0)$-theory and $(\theta,\theta)$-theory as long as the operator which gives the quantum number in $(\theta,\theta)$-theory is given by replacing all the $\phi^0$ and the products of the corresponding operator in $(0,0)$-theory\footnote{For example, this is the case for the momentum operator. Indeed, if we replace all the $\phi^0$ and the products in the definition of $\mathcal{P}^\mu$ (eq.(\ref{old momentum})) by $\phi^\theta$ and the star product, the resulting operator turns out to be equal to $\mathcal{P}_\theta^\mu$ by ignoring total derivative terms in the integrand.}.

	With these correspondences for field operators and Hilbert spaces, it is easily found that matrix elements of $\mathcal{O}_\star[\phi^\theta]$ in $(\theta,\theta)$-theory are also equal to those of $\mathcal{O}[\phi^0]$ in $(0,0)$-theory.
	This follows from eq.(\ref{star product s-s}) and eq.(\ref{two actions}) immediately:
\begin{equation}
	\langle\alpha|\star\mathcal{O}_\star[\phi^\theta]\star|\beta\rangle=\langle\alpha|\Big(\mathcal{O}_\star[\phi^\theta]\star|\beta\rangle\Big)=\langle\alpha|\mathcal{O}[\phi^0]|\beta\rangle.\label{map of matrix elements}
\end{equation}
	This implies that the vacuum expectation values of time-ordered products of fields for two theories are equal each other:
\begin{equation}
	\langle0|T_\star\phi^\theta(x_1)\cdots\phi^\theta(x_n)|0\rangle=\langle0|T\phi^0(x_1)\cdots\phi^0(x_n)|0\rangle.
\end{equation}
	From this we can find that the Green's function $G^{(n)}_\star$ in eq.(\ref{green}) are equal to the ordinary Green's functions $G^{(n)}$.
	This is the fact which we have found by the perturbation expansion in section 4.1.

	Eq.(\ref{map of matrix elements}) also means that n-fold vacuum expectation values of field operators, i.e., Wightman functions for these theories are the same:
\begin{equation}
\begin{array}{rcll}
	\mathcal{W}_\star(x_1,\cdots,x_n)&=&\langle 0|\star\phi^\theta(x_1)\star\cdots\star\phi^\theta(x_n)\star|0\rangle\\
	&=&\langle 0|\phi^0(x_1)\cdots\phi^0(x_n)|0\rangle=:\mathcal{W}(x_1,\cdots,x_n).\rule{0mm}{7mm}
\end{array}
\end{equation}
	It is suggestive to see this $\mathcal{W}_\star$ has the same form as the Wightman function adopted in \cite{Chaichian:2004qk}.
	Moreover, $\mathcal{W}_\star$ satisfies all the axioms proposed in \cite{Chaichian:2004qk}.
	In particular, the axiom of the local commutativity is resulted from the star-deformed commutation relation eq.(\ref{nccr0}).
	That is, our new quantization provides a specific field theory, $(\theta,\theta)$-theory, which satisfies all the axioms in \cite{Chaichian:2004qk}.

	It is straightforward to extend the discussion here to arbitrary $(\theta, \tilde{\theta})$-theory.
	We can establish the correspondence between $(\theta, \tilde\theta)$-theory and $(0, \Theta)$-theory by the aid of the map
\begin{equation}
	\phi^\theta(x)=\exp\left(-\frac{1}{2}\mathcal{P}_\Theta\theta\partial\right)\phi^0(x),
\end{equation}
	where $\phi^\theta(x)$ and $\phi^0(x)$ give field operators of $(\theta,\tilde\theta)$-theory and $(0,\Theta)$-theory respectively, and $\mathcal{P}_\Theta^\mu$ is a momentum operator of $(0,\Theta)$-theory.

	Finally, as an example, let us apply the map (\ref{theory map}) to the free field theory.
	In the case of the free field, this map gives the correspondence between creation and annihilation operators, and Fock spaces of two theories.
	First, for creation and annihilation operators, the map (\ref{theory map}) implies
\begin{equation}
	a_\bk^\theta=\exp\left(\frac{i}{2}\mathcal{P}\theta k\right)a_\bk^0,\quad {a_\bk^\theta}^\dagger=\exp\left(-\frac{i}{2}\mathcal{P}\theta k\right){a_\bk^0}^\dagger,\label{map for ladder}
\end{equation}
	where $a_\bk^\theta$ and ${a_\bk^\theta}^\dagger$ are creation and annihilation operators of the free scalar field quantized by the star-deformed commutation relation, i.e., $a_\bk$ and ${a_\bk}^\dagger$ in section 3, and $a_\bk^0$ and ${a_\bk^0}^\dagger$ are those quantized by the ordinary canonical quantization, i.e., the standard creation and annihilation operators.
	This correspondence is the same one which appears in \cite{Balachandran:2005eb}.

	For the correspondence between Fock spaces, we firstly notice that we can use the same state as a vacuum state for two theories.
	This fact is easily seen by the following relation which results from eq.(\ref{map for ladder}),
\begin{equation}
	a_\bk^0|0\rangle=0\Rightarrow a_\bk^\theta|0\rangle=0.
\end{equation}
	Based on this relation and eq.(\ref{map for ladder}), we easily find
\begin{equation}
	a_{\bk_1}^{\theta\dagger}\star\cdots\star a_{\bk_n}^{\theta\dagger}|0\rangle=a_{\bk_1}^{0\dagger}\cdots a_{\bk_n}^{0\dagger}|0\rangle.
\end{equation}
	This equation means the star-deformed Fock states we argued in section 3 are equal to the ordinary Fock states.

\vspace{9mm}

\section{Conclusion and Discussion}\mbox{}

	In this paper, we have presented a new quantization scheme for noncommutative field theories, where we promote fields to operators by imposing the star-deformed canonical commutation relation (\ref{nccr0}) instead of the standard canonical commutation relation.
	Based on this deformed commutation relation, a consistent quantum theory is constructed, e.g., the Fock space representation (\ref{Fock state}) for the free field provides a physical picture of a particle.
	We find, unexpectedly, that the resulting dynamics are the same as the commutative field theory, even though the classical action possesses a noncommutative interaction term.
	This is caused by the cancellation between the noncommutativity in the interaction term and the noncommutativity introduced in the quantization.
	We are thus allowed to apply the renormalization procedure in the parturbation calculation in exactly the same way as the ordinary commutative QFT.

	The noncommutativity cancellation observed here does not mean that the new quantization is trivial.
	Indeed, further observations of the structure of this noncommutativity cancellation show that, if we use different noncommutative parameters for the interaction term and the quantization (i.e., $(\theta,\tilde\theta)$-theory), the resulting dynamics are the same as those of the noncommutative QFT with noncommutative parameter $\Theta^{ij}=\tilde\theta^{ij}-\theta^{ij}$.
	In particular, starting from a commutative action and quantizing fields by the noncommutative quantization (i.e., $(-\theta,0)$-theory), one obtains a theory equivalent to the ordinary noncommutative QFT with the noncommutative interaction term and fields quantized by the canonical quantization (i.e., $(0,\theta)$-theory).
	This indicates that we can also introduce the noncommutativity of a noncommutative QFT in the quantization procedure rahter than in the action as conventionally done.
	It may be possible to make use of this structure to study the ordinary noncommutative QFT using the canonical quantization.
	On the other hand, this structure implies that it is impossible to tell which of the quantizations, the canonical quantization or the deformed one, is realized in nature from any experimental observations.
	This is because all the $(\theta,\tilde\theta)$-theories with the same value of $\Theta^{ij}=\tilde\theta^{ij}-\theta^{ij}$ yield the same dynamics.
	By observations, we can only determine the value of $\Theta^{ij}$.

	Finally, we mention another possibility for the Fock space representation.
	Instead of the Fock space representation defined in section 2.3, we could also define the basis vectors as
\begin{equation}
	|\bk_1,\bk_2,\cdots,\bk_n)=a_{\bk_1}^\dagger a_{\bk_2}^\dagger\cdots a_{\bk_n}^\dagger|0\rangle ,
\end{equation}
	in place of eq.(\ref{Fock state}).
	These basis vectors also provide energy-momentum eigenstates on account of eq.(\ref{momentum commutation relation}).
	The difference from the original one eq.(\ref{Fock state}) arises only in the phase factor:
\begin{equation}
	|\bk_1,\bk_2,\cdots,\bk_n\rangle=\exp\left[-\frac{i}{2}\sum_{i<j}^nk^i\theta k^j\right]|\bk_1,\bk_2,\cdots,\bk_n).
\end{equation}
	We can immediately find that the difference results in a change of the statistics of a particle.
	In fact, if we consider $|\bk_1,\bk_2,\cdots,\bk_n)$ as particle states, those particles obey nontrivial statistics due to the momentum dependent phase factor, rather than the conventional Bose statistics.

	Although the quantization we present here is only one of many possible quantizations, we believe that our results should be sufficient to suggest that the canonical quantization is not the only viable quantization for noncommutative field theories.

\vspace{1cm}

\noindent{\bf \large{Note added}}\mbox{}\newline

	After this paper was posted on the hep-th archives, we have been informed that \cite{Oeckl:1999zu,Oeckl:2000eg} include results related to this paper.
	In particular, the Green's functions considered in \cite{Oeckl:2000eg} can be constructed from the field operator quantized by the noncommutative quantization in this paper.
	Indeed, if we define Green's functions without star products rather than (\ref{green}), they would agree with those of \cite{Oeckl:2000eg}.
	In addition, the equivalence of Green's functions of a QFT and its twisted version demonstrated perturbatively in \cite{Oeckl:2000eg} is essentially the same as the noncommutativity cancellation proved even nonperturbatively in this paper.
	These facts imply that \cite{Oeckl:1999zu, Oeckl:2000eg} could provide the path integral formulation for the noncommutative quantization in this paper.

\vspace{1cm}

\noindent{\bf \large{Acknowledgments}}\mbox{}\newline

	I would like to thank Izumi Tsutsui for many useful discussions and a careful reading of the manuscript.
	I am also grateful to Rabin Banerjee for valuable discussions and remarks.\newline\newline

\end{document}